\documentclass[pdflatex,sn-mathphys-ay]{sn-jnl}
\usepackage{booktabs} 
\usepackage{amsmath}
\usepackage{amsfonts}

\usepackage{multirow}
\usepackage{array}
\usepackage{longtable}

\usepackage{gensymb}

\usepackage{natbib}

\usepackage{changepage}
\usepackage{hyperref}
\usepackage{graphicx}

\setcitestyle{authoryear}

\defcitealias{sandholm1996multiagenta}{S96}
\defcitealias{waltman2008qlearning}{WK08}
\defcitealias{wunder2010classesb}{W10}
\defcitealias{calvano2020artificiala}{C20}
\defcitealias{eschenbaum2021robust}{E21}
\defcitealias{hansen2021frontiers}{H21}
\defcitealias{klein2021autonomousa}{K21}
\defcitealias{asker2022artificial}{A22}
\defcitealias{sanchez-cartas2022artificial}{SCK22}
\defcitealias{abada2023artificial}{AL23}
\defcitealias{banchio2023adaptive}{BM23}
\defcitealias{leonardos2023catastropheb}{L23}
\defcitealias{hartline2024regulation}{H24}
\defcitealias{carissimo2024counterintuitive}{C24}

\usepackage[ruled]{algorithm2e}

\SetAlFnt{\small}
\SetAlCapFnt{\small}
\SetAlCapNameFnt{\small}
\SetAlCapHSkip{0pt}
\IncMargin{-\parindent}

\newtheorem{prop}{Result}

\newtheorem{proposal}{Proposal}
\newtheorem{scene}{Scenario}

\begin{document}

\title{Algorithmic Collusion is Algorithm Orchestration}

\author*[1]{\fnm{Cesare} \sur{Carissimo}}\email{ccarissimo@ethz.ch}

\author[2]{\fnm{Fryderyk} \sur{Falniowski}}\email{falniowf@uek.krakow.pl}

\author[3]{\fnm{Siavash} \sur{Rahimi}}\email{srahimis@caltech.edu}

\author[4]{\fnm{Heinrich H.} \sur{Nax}}\email{hnax@ethz.ch}

\affil*[1]{\orgdiv{Computational Social Science}, \orgname{Swiss Federal Institute of Technology Zurich}, \orgaddress{\street{Stampfenbachstrasse}, \city{Zurich}, \postcode{8006}, \country{Switzerland}}}

\affil[2]{\orgdiv{Krakow University of Economics}, \orgaddress{\street{Rakowicka 27}, \city{Kraków}, \postcode{31-510}, \country{Poland}}}

\affil[3]{\orgdiv{California Institute of Technology}, \city{Pasadena}, \postcode{CA 19925}, \country{USA}}

\affil[4]{\orgdiv{Behavioral Game Theory}, \orgname{University of Zurich}, \orgaddress{\street{Andreasstrasse}, \city{Zurich}, \postcode{8050}, \country{Switzerland}}}

\keywords{Algorithmic Collusion, Bertrand Duopoly, Meta Game, Q-Learning.}

\abstract{We propose a fresh `meta-game' perspective on the problem of algorithmic collusion in pricing games a la Bertrand. Economists have interpreted the fact that algorithms can learn to price collusively as tacit collusion. We argue instead that the co-parametrization of algorithms, in ways as are necessary to obtain algorithmic collusion, typically requires algorithm designers to engage in some form of explicit collusion or `algorithm orchestration.' In our model, the algorithm designers play a meta-game of parametrizing their algorithms, which then play repeated Bertrand competition. The strategic analysis at the meta-level reveals new equilibrium and collusion phenomena. (JEL: C62, C63, D43, L13)}

\maketitle

\section{Introduction}

Algorithmic collusion is a growing phenomenon and a major challenge for the law and regulators \citep{schwalbe2018algorithms}. Already prior to the advent of AI and the rise of pricing algorithms, because one cannot `read the minds of managers,' collusion (without algorithms) was notoriously difficult to detect. Now, with the rise of relevant technologues, detection and regulation become even more difficult as pricing algorithms operate faster and by themselves and thus operate in legal grey zones. In fact, the law in most countries is such that not all behaviors resulting in collusion are deemed unlawful. In particular, it is not generally unlawful for price setters to sustain high prices through self-enforcing pricing strategies with so-called `reward-punishment schemes,' where competitors tacitly recognize the reward of supra-competitive prices as well as the threat of punishment via low prices. The current interpretation of antitrust in many countries deems collusion as unlawful only when there is \textit{explicit} communication between price setters. 
In legal language, \textit{an agreement among competitors to limit competition is unlawful} but \textit{conscious parallelism} is not as long as it is competitively arrived at, \textit{[...] not because such pricing is desirable (it is not), but because it is close to impossible to devise a judicially enforceable remedy for `interdependent' pricing}  \citep{harrington2018developingb}.\footnote{From Clamp-All Corporation versus Cast Iron Soil Pipe, 851 F.2d 478 (1st Cir. 1988).}

An emerging economic literature studying algorithmic collusion is interpreted as having shown that the conditions for \emph{tacit} collusion, which would not be deemed illegal, are increasingly met as algorithms are able to communicate indirectly through actions without ever involving human representatives of the competitors directly and communicating in a traditional sense. We shall make the point that adjudicating algorithmic collusion as tacit or explicit requires taking a fresh look at the problem, where the co-parameterization of the algorithms themselves is also investigated.

\textbf{Our perspective is that} \textit{a strategic analysis of the individual algorithms' designs ought to be conducted in order to establish whether the interdependent algorithms are designed competitively, that is, as best responses to one another seeking higher payoff unilaterally, or non-competitively, even collusively, so as to increase payoffs jointly.}

\textbf{Our main finding is that} \textit{algorithmic collusion is typically achievable only if algorithms are orchestrated, that is, jointly trained and parameterized non-competitively in the pursuit of supra-competitive prices and joint payoff improvements. By contrast, algorithms set competitively will end up pricing close to competitively.}

We believe that our perspective provides a timely alternative to the existing criteria according to which algorithmic collusion is evaluated, especially as internet platforms increasingly distribute their custom pricing algorithms to their market participants. Evidently, there is a massive conflict of interest for the platform as the algorithm orchestrator in deploying collusive algorithms that curate the entire price landscape jointly.
Indeed, this conflict of interest has been acknowledged in very recent applied legal scholarship \citep{harrington2025hub}. In these `Hub and Spoke' style situations where a third-party has full control, pricing algorithms can be orchestrated to learn the most collusive strategies. It remains an open question how easily pricing algorithms will \textit{tacitly} collude without the need for algorithm orchestration.

Our framework addresses the distinction between legal and illegal collusion in a novel way that we believe is necessary for regulation to be welfare-enhancing: whether the interdependent algorithms are competitively parametrized or the result of algorithm orchestration. We bring our understanding of collusion to bear on the most popular class of reinforcement learning algorithms in the literature on algorithmic collusion. Indeed, we show that out-ruling these algorithms altogether may not benefit consumer surplus much, as algorithms that are parametrized as best responses to one another would behave competitively in the pricing game. By contrast, we find that the kind of highly collusive pricing behavior as previously identified in the literature would require explicit collusion in parameter setting by the algorithm designers.

We obtain our results embedding the algorithmic interaction of competing algorithms in a `meta-game' of algorithm design that is played by competing firms. You may think of this meta-game as played by algorithm designers who pick the (hyper)parameters\footnote{A hyperparameter that \textit{determines how} an algorithm learns must be distinguished from the `weights' (sometimes called parameters) that the algorithm updates during learning.} of their pricing algorithms, thus affecting both their own and the other parties’ payoffs.
Importantly, we postulate that the relevant players playing the meta-game (i.e. the algorithm designers) must tune their algorithm against their algorithmic opponents seeking to maximize their payoffs unilaterally. To the best of our knowledge, our extensive computational analysis is novel in economics, yet deemed relevant in principle by legal scholars who recognize the dynamic improvement process of `The Algorithm Game' \citep{bambauer2018algorithm}, which was a source of inspiration. 

The remainder of this paper is structured as follows. First, \autoref{sec:background} is dedicated to a review of the literature on algorithmic collusion. In \autoref{sec:model}, we formalize the meta-game and show how we compute the meta-game payoffs. In \autoref{sec:best-repsonses} we take the perspective of competing algorithm designers and characterize the meta-game Nash equilibria. In \autoref{sec:collusion} we discuss collusion in the meta-game, defining co-training, co-parameterization, and tacit collusion. Finally, in \autoref{sec:conclusion}, we discuss avenues for future work and conclude with a proposal of a simple test to differentiate between collusive and competitive algorithms by analysis of emergent price patterns.

\section{Related Literature}\label{sec:background}

\subsection{Q-learning in Bertrand duopoly}

A growing body of economic literature studies learning in games of competition (a la Bertrand, Cournot, etc.) and other games (e.g. Prisoner's dilemma games) as played by interacting algorithms. Several papers identify situations where learning algorithms converge to collusive outcomes that yield supra-competitive payoffs\footnote{The present paper and the literature we discuss define algorithms as reinforcement learning agents that are simulated in computational experiments. A related line of theoretical results finds that algorithms, defined as fixed price mappings that are updated asynchronously to best respond to an opponents fixed price mapping, inevitably lead to near monopoly profits \citep{salcedo2015pricing, lamba2022pricing}.}. In this section, with a focus on Q-learning in Bertrand duopoly games, we review the relevant design choices made in the literature and relate them to the kinds of convergent phenomena that have been observed. We organize the existing literature according to the following two dimensions along which existing studies differ:
\begin{description}
    \item[(i)] the learning algorithms being used---algorithms differ according to how they evaluate payoffs and how they select among available actions. Much of the literature and indeed the one we shall focus on and contribute to is concerned with Q-learning algorithms that vary in terms of their hyperparameters, especially concerning exploration (exp\_), being $\epsilon$-greedy ($\epsilon$-g), where the rate $\epsilon$ can be decayed or fixed during learning. We also distinguish papers by whether they allow agent asymmetry (asym\_).  
\item[(ii)] how the chosen algorithm is being evaluated---broadly, one can distinguish between cumulative (online) and converged (limit) evaluations, the choice of which (eval\_) determines the optimal balance of exploration and exploitation; an algorithm that is evaluated after convergence is free to explore without repercussion during learning, while exploration for an algorithm that is evaluated cumulatively is costly.
\end{description}

\begin{longtable}{l|l|l|l|l|l}

\caption{Q-learning in Bertrand duopolies.}
\label{tab:meta_game_literature}\\
\toprule
             Paper &                            exp\_ &    asym\_ &                       eval\_ & observed pricing phenomena & meta-Nash \\
\midrule
\endfirsthead
\midrule
\endhead
\midrule
\multicolumn{6}{r}{{Continued on next page}} \\
\midrule
\endfoot

\bottomrule
\endlastfoot
   
    \citetalias{calvano2020artificiala} &        decay &            no &        limit &  \textit{focal} & no \\
    \citetalias{eschenbaum2021robust} &          decay &           yes &        limit &  out of sample: \textit{reduced focal}  & no \\ 
    \citetalias{klein2021autonomousa} &          decay &            no &        limit &  turn taking: \textit{Edgeworth cycles} & no \\
    \citetalias{asker2022artificial} &           none &            no &       online &  async and memory-less: \textit{focal} & no \\ \hline
    ours &           fixed &           yes &       online & \textit{noisy competitive} & yes \\ \hline
\end{longtable}

\autoref{tab:meta_game_literature} summarizes the existing literature on $Q$-learning in Bertrand duopoly games. Some of the more prominent studies have shown that price collusion is possible between algorithms. Focal price collusion at high prices can be achieved by $Q$-learners that are symmetrically parameterized, decay their exploration rates, and train and deploy with their opponents and are evaluated after convergence \citep{calvano2020artificiala}. With a similar setup, Edgeworth cycles are learned in if players take turns changing prices \citep{klein2021autonomousa}. The work of \cite{eschenbaum2021robust} shows that the `clean’ collusive outcomes achieved by the aforementioned setups break down when algorithms face new opponents out of training, and are replaced by competitive pricing, evidencing the need for coordination in training and deployment for algorithmic collusion to emerge. Indeed, it is also shown that algorithmic collusion is less likely to emerge when $Q$-learning updates can use counterfactuals from downward sloping demand to update unobserved pricing outcomes \citep{asker2022artificial}.\footnote{The literature on algorithmic collusion concerned with games beyond the Bertrand duopoly, and with algorithms beyond $Q$-learning is discussed in Appendix A.} 

In the wider literature, online Q-learning is shown to create noisy coordination in the Prisoners Dilemma \citep{wunder2010classes}, and in Braess's Paradox \citep{carissimo2024counterintuitive}, but when $Q$-learning is analyzed in the limit, it is proven to converge to cooperation in the Prisoner's Dilemma \citep{dolgopolov2024reinforcement}, and in the case of spontaneous coupling \citep{banchio2023adaptive}, for a narrow range of parameters. In general, it has been most common to evaluate $Q$-learning in the limit; in Cournot oligopoly by \citet{waltman2008qlearning}, in Bertrand duopoly by \citet{calvano2020artificiala, klein2021autonomousa, eschenbaum2021robust}, and in pricing games with storage capacity by \citet{abada2023artificial}. This means that the $Q$-learners are free to explore during learning, which means that their exploration is some kind of `costless pre-game.' 

The designs of exploration policies vary between $\epsilon$-greedy policies \citep{calvano2020artificiala, klein2021autonomousa, eschenbaum2021robust, abada2023artificial}, Boltzmann policies \citep{sandholm1996multiagenta, waltman2008qlearning, leonardos2023catastropheb}, and upper confidence bound algorithms \citep{hansen2021frontiers}. Independent of exact implementation, these policies often handle the exploration-exploitation trade-off similarly: the learner's price setting is random in the initial learning phase, and the randometa-NEss decreases such that the learner's price setting eventually becomes deterministic and price fluctuations are minimized.

One thing that is important to note is that most contributions in the literature focus on interactions where all algorithms are parametrized symmetrically. There are some exceptions to this pattern. The aforementioned work of \citet{sandholm1996multiagenta} tested $Q$-learning against fixed strategies, and \citet{sanchez-cartas2022artificial} creates high asymmetry by testing $Q$-learners against Particle Swarm Optimization. The work of \citet{eschenbaum2021robust} experiments with some heterogeneity, based on the evaluation of $Q$ learners against each other without first training with each other. The work of \cite{denboer2024artificial} remarks on the importance of asymmetric algorithmic strategies and mentions the `meta-game in which firms choose algorithms in order to enhance their profit'. Finally, the work of \citet{brown2023competition} tests pricing strategies that vary in their frequency of price adjustments finds that `asymmetric pricing technology can increase price levels, exacerbate the price effects of mergers, and generate price dispersion'.

\subsection{Us vis-à-vis the Literature }

We have learned from the literature that price collusion between two (and more) learning algorithms is possible. In particular, it is shown to be possible in certain ranges of the parameter spaces of the learning algorithms. What is not known from the literature is whether algorithmic collusion would naturally arise from algorithms being designed by competitors or by allies. This motivates our main research question, which to us seems highly appropriate and timely as regulators strive to adequately regulate algorithmic collusion:

\vspace{0.5cm}

\noindent\textbf{RQ:} Does price collusion require algorithm orchestration?
    \begin{enumerate}
        \item What pricing dynamics emerge among competitive algorithms?
        \item Instead, what are the pricing dynamics of orchestrated algorithms?
    \end{enumerate}

Our focus shifts away from the phenomenon of `algorithmic collusion' to the phenomenon of collusive algorithm parametrization, which we shall refer to as \textit{algorithm orchestration}. To make this perspective explicit, we study the incentives of algorithm designers by formulating a meta-game where the strategies of the players, that is, of the algorithm designers are the hyperparameters that they pick for their algorithms that then compete in an underlying repeated game. We thus distinguish between collusion in the meta-game and price collusion in the underlying repeated game. While we focus on the kinds of algorithms and games that have been studied in the literature, this conceptual framework can be applied to any combination of algorithms and underlying game. We tackle concretely the most popular setting, which is a Bertrand duopoly game played by two $\epsilon$-greedy $Q$-learners.
We find that the meta-game has pure Nash Equilibria, and algorithms parametrized at the Nash Equilibrium exhibit pricing behaviors that are only \textit{slightly} supra-competitive and noisy. By contrast, more collusive pricing behaviors are obtained by non-competitive play of the meta-game which can lead at best to the Pareto front, which we find to constituted of asymmetric parameter combinations.

\section{Algorithm Competition Versus Algorithm Orchestration}\label{sec:model}

In order to determine which algorithms are designed competitively and which algorithms are designed cooperatively (orchestrated) we must define the design choices, and the outcomes of algorithmic interaction: what we call a 'meta-game'. In the previous section we identified two important design dimensions in the literature on $Q$-learning in the Bertrand Duopoly, namely how the algorithm picks exploratory actions and how the algorithm is evaluated. Other design choices are how quickly the algorithm should respond to fluctuations, how much to discount profits in the future, the choice of the minimum and maximum prices, how far into the past to consider opponent prices, and how to model the opponent, to name a few. If we take the game that ensues from all possible combinations of all possible designs, and call this the master-meta-game, it is clear that this game is enormous and may be infinite. Nevertheless, we can get valuable insights by understanding the meta-incentives at play where learning algorithms are deployed in market competition. To initiate this exploration, we propose to continue in line with recent work on algorithmic competition and restrict ourselves to a meaningful and representative sub-game of the master-meta-game, comprised of $Q$-learners that compete in a repeated Bertrand Duopoly. In this section, we will formalize the meta-game and formulate the specific meta-game that we evaluate in our computational experiments.

\subsection{The meta-game}\label{sec:meta-game}

The players in the meta-game are the algorithm designers, each of which control one algorithm and set the hyperparameters of their algorithm. The player's payoffs are determined by competition of their algorithms in a repeated game, which in this paper is the repeated Bertrand duopoly. Algorithms map past prices (and possibly other relevant information) to the next prices to be set. For the case of two players, `ego’ and `alter,’ each player chooses a parameter (or vector of parameters) for their algorithm, and an algorithm is fully characterized by its parameters. We represent the parameter space as $\Theta$ and grant both players access to the same space. We denote a player's choice of parameter as $\theta_i \in \Theta$.

The payoffs for ego and alter are determined by evaluating the performance of their algorithms in repeated play of the Bertrand Duopoly. The game is repeated for $T$ rounds. At each round ego and alter's algorithms are queried to determine their prices for that round. The demand in each round is a linear decreasing function of the price. The reward in each round is the round's profits, calculated as their share of the demand multiplied by the price. If the algorithms set the same price, demand is split equally. Otherwise, the lowest price attracts the entire demand. At the end of the $T$ rounds, both algorithms will have played a vector of prices $\Vec{p}_i=(p_1, \dots, p_T)_i$ and experienced a vector of profits $\Vec{r}_i=(r_1, \dots, r_T)_i$.

The payoffs for the designers of the algorithms are dependent on the parameters as a function $\pi: \Theta \times \Theta \rightarrow \mathbb{R} \times \mathbb{R}$; a parameter profile $\theta_1$, $\theta_2$ is assigned to a payoff for both players, with $\pi(\theta_1, \theta_2) = (\pi_1, \pi_2)$. Implicitly, the payoffs for the designers of the algorithms are also dependent on the underlying game where the algorithms are evaluated against each other, which in our case is the repeated Bertrand Duopoly.

If alter picks a parameter $\theta_2$, the best response parameter $\theta^*$ of ego is determined with the best response function $br: \Theta \rightarrow \Theta$. Specifically, the best response parameter is $\theta^*$ such that for all $\theta \in \Theta\backslash\{\theta^*\}$, $\pi(\theta^*, \theta_2)_1 > \pi(\theta, \theta_2)_1$, where $\pi(\theta, \theta_2)_1$ is the payoff for the first player picking $\theta$ against $\theta_2$. Then, a pure Nash equilibrium of the meta-game is any parameter profile $(\theta_1, \theta_2)$ that is a mutual best response in the parameter picking meta-game: $br(\theta_1)=\theta_2$ and $br(\theta_2)=\theta_1$.

The Nash Equilibria of the meta-game are what we consider \textit{algorithmic competition}. However, the algorithm designers that play the meta-game may also collaborate, what we call \textit{algorithmic orchestration}. In the case of orchestration, the players may instead reach the Pareto frontier of the meta-game, which is the set of all parameter profiles $(\theta_1, \theta_2)$ such that there is no $\theta' \in \Theta$ where $\pi(\theta', \theta_2)_1 > \pi(\theta_1, \theta_2)_1$ and $\pi(\theta', \theta_2)_2 \geq \pi(\theta_1, \theta_2)_2$.\footnote{By symmetry of the meta-game, and the underlying Bertrand Duopoly, this condition expressed solely for one player is sufficient to determine the entire Pareto front.}

\subsection{Parameters of the meta-game}\label{sec:methods}

In the previous section we formalized the meta-game, and in this section we formulate our specific implementation of the meta-game, on a high level, justifying our restrictions of the algorithmic design space. In the next section we will describe our detailed computational setup.

We begin our formulation by fixing the meta-game to players that parametrize their algorithms in the class of $Q$-learning algorithms, and deploy their algorithm against the other players algorithm in a repeated Bertrand duopoly. We take inspiration from the setups of \citet{calvano2020artificiala, klein2021autonomousa, eschenbaum2021robust, asker2022artificial}, and reflect on their particular design choices.

We seek a meaningful and representative formulation of the meta-game of algorithm design to answer our main research question: `does price collusion require algorithm orchestration?' In the language of the meta-game, this question can be reformulated as: `what are the parameters for $Q$-learning algorithms that algorithm designers should pick to unilaterally maximize their payoff?' Since the space of all design parameters for $Q$-learning algorithms is large, we also wish to restrict the design space $\Theta$ to a representative subspace.

We decide to exclude design setups akin to centralized training (co-training for short), commonly known in the computer science literature as \textit{centralized training and decentralized execution} (CTDE), which is designed to improve the coordination between multi-agent systems of reinforcement learners. CTDE separates the algorithmic process into two phases. The first phase is a training phase where the algorithms can coordinate with each other -- more than they would be able to otherwise during deployment.\footnote{For example, designers can coordinate their algorithms to collaboratively anneal their learning parameters (namely learning rate $\alpha$ and exploration rate $\epsilon$ for $Q$-learning), share useful payoff information (gradients) and engage in costless joint exploration.}
In the second phase following centralized training, the decentralized execution phase, learning stops (which means that the reinforcement learning algorithms no longer use new information to update their behavior), and the algorithms are deployed and evaluated with each other. Therefore, this second phase is akin to what we termed `limit' evaluations in \autoref{sec:background}. 

Setups with CTDE enable substantial coordination between learning agents, by design. We argue that they also require explicit coordination between the algorithm designers. As such, they are not unilaterally competitive designs of the meta-game. Instead, they represent \textit{collusion} in the meta-game.
 
Therefore, we exclude two setups: coordinated choices of learning parameters, and evaluations after convergence (in the limit). As such, we formulate a meta-game with two key features:

\begin{enumerate}
    \item[a.] algorithm designers choose independently the learning parameters for $Q$-learning.
    \item[b.] algorithm designers evaluate the performance of $Q$-learning \textit{online}, meaning that the profits achieved during the learning/training of the $Q$-learning algorithms will be factored into the payoff of the designers. 
\end{enumerate}

In other words, the intuition behind the formulation of our meta-game is that the algorithms are \textit{always learning}, and the algorithm designers have control over their algorithm's learning parameters. In practice, we break down this dynamic setting into a stage game, where algorithm designers pick the parameters of their algorithms, and their algorithms are then tested against each other. In the next section we describe the computational setup of our meta-game in detail.

\subsection{The computational setup}

In this section we describe the details for setting up our meta-game. We cover the $Q$-learning algorithm, the strategy space for the meta-game players, and the payoffs for the meta-game players. In the previous section we argued why our meta-game involves designers picking learning parameters for $Q$-learning algorithms that are evaluated against each other \textit{online}. 

First, we are going to describe learning parameters which 
are all the design choices available to the players in our meta-game. In the following sections we will describe 
other choices that characterize our specific meta-game: the number of price levels, the length of $Q$-learner's memory, the $\epsilon$-greedy exploration policy, the discretization of the strategy space, the number of repeated game rounds, the uniform weighted average for their payoff, and the initialization of the $q$-values. The extent to which the variation of this setup influences meta-game incentives is left to future work.

\subsubsection{The Q-Learning algorithm}

$Q$-learning is used widely to study learning, and has been center and focus of the algorithmic collusion literature. Optimal in Markov Decision Processes (MDP), $Q$-learning is guaranteed to reach an optimal deterministic policy if exhaustive exploration is allowed \citep{watkins1992q}. An MDP is defined by a state space $\mathcal{S}$, an action space $\mathcal{A}$, a reward function $R: \mathcal{S} \times \mathcal{A} \rightarrow \mathbb{R}$, and a transition function $\mathcal{T}: \mathcal{S} \times \mathcal{A} \times \mathcal{S} \rightarrow [0,1]$. However, in a multi-agent setting $Q$-learning does not have convergence guarantees.

In our setup, two $Q$-learners compete in a pricing game. The action space of their pricing game is a set of discrete prices in a range $\mathcal{A}=\{0,1,...,6\}$. They perceive the last price of their opponent and only condition their next action on this information, so the state space is $\mathcal{S}=\{0,1,...,6\}$. Then, they select the next prices simultaneously. 

We use a tabular implementation of $Q$-learning so all $q$-values are stored in a table. The Q-table $Q: \mathcal{S}\times \mathcal{A} \rightarrow \mathbb{R}$ maps states and actions to the cumulative sum of discounted future rewards.

$Q$-learning has three parameters that determine: the learning rate $\alpha \in (0,1]$, the exploration rate $\epsilon \in [0,1]$, and the discount factor $\gamma \in [0,1]$. 
The learning rate governs how quickly $q$-values change given a new transition from state $s$ to state $s'$ with action $a$ receiving a reward $r$, following the Bellman update rule:
\[Q(s,a) \leftarrow Q(s,a) + \alpha (r + \gamma \max_{a'}Q(s',a') - Q(s,a)).\]

In the Bellman update appears the discount factor $\gamma$ which determines how much actions in the future influence the present. The exploration rate governs the likelihood of picking random actions, exploratory actions, following an $\epsilon$-greedy action policy from a state $s$:

$$\epsilon\text{-greedy}(s) = \begin{cases} arg\max_a Q(s,a) \:\: & w.p. \: 1 - \epsilon \\ \text{uniform random}\{\mathcal{A}\} \:\: & w.p. \: \epsilon \end{cases}.$$

\subsubsection{The strategy space}
At the start of the meta-game each player picks a tuple of parameters $(\alpha, \epsilon, \gamma)_i = \theta_i \in \Theta = [0,1]^{\times3}$. Each parameter admits values between $0$ and $1$. We discretize the parameter space such that each parameter is allowed $10$ equidistant values between the following natural specified endpoints: $0.01 \leq \alpha \leq 1$; $0 \leq \epsilon \leq 0.5$; $0 \leq \gamma \leq 1$. We set these bounds in between the reasonable extreme policies. Note that we do not include $\alpha=0$ as this would negate all $q$-value updates. Moreover, we do not consider $\epsilon$ higher than $0.5$, because when $\epsilon>0.5$ the role of learning is overwhelmed by randomness, 
and we believe that higher values are not realistic choices for the exploration rate of an online $Q$-learning algorithm interested in maximizing the cumulative sum of discounted future rewards, particularly given the focus of the prior literature on decayed exploration. 

\subsubsection{The repeated game payoffs}\label{sec:method-payoff}

As is common for Bertrand Duopolies, consumers are price inelastic and always choose the producer with the lowest price. When the prices are equal, they choose one producer randomly. Demand varies linearly based on the price, that is $D_t= max_{a\in\mathcal{A}} a - min\{p_{1,t},p_{2,t}\}$. There is no marginal cost of production.

After both algorithm designers pick parameters for their $Q$-learning algorithm $\theta_1$ and $\theta_2$, the two algorithms are deployed with their parametrization to learn in a Bertrand Duopoly where they play $T=40000$ rounds. The payoff of a parameter profile $\theta_1, \theta_2$ is expressed as $\pi(\theta_1, \theta_2)$ and is determined by repeated play of the Bertrand Duopoly. In each round $t$ player $i$ receives a profit $r_{i,t}$. Each player accumulates a vector of profits $\mathbf{r}=(r_1, \dots, r_{T})$ over the course of repeated play of the Bertrand Duopoly. The average of this vector is $\hat{\mathbf{r}} = \frac 1T \sum_{t=1}^{T}r_t$. The estimated average profits for a parameter profile $(\theta_1, \theta_2)$ are $\pi(\theta_1, \theta_2)=(\hat{\mathbf{r}}_1, \hat{\mathbf{r}}_2)$. Finally, the payoffs for each player of the meta-game will be the online performance of their algorithm expressed as the average profit $\hat{\mathbf{r}}$ over the learning period $T$.


We simulate every parameter profile $(\theta_1, \theta_2) \in \Theta^2$ to estimate the payoffs $\pi(\theta_1, \theta_2)$. For $3$ learning parameters that take $10$ possible values, we have $1000$ parameter choices per player and $1$ million parameter profiles. Each parameter profile is evaluated $40$ times for statistical averaging.\footnote{We choose these parameters as a compromise between computation time and convergence issues.} 
The $q$-values of the $q$-tables for each $Q$-learning algorithm are initialized as follows:

\[Q(s,a_i) = \frac{\sum_{a_{-i} \in \mathcal{A}} \pi(a_i, a_{-i})}{(1-\gamma)|\mathcal{A}|}.\]
which is the value they would converge to if their opponent played a uniform random pricing strategy.

We report the pricing outcomes in terms of profit gain as measured by \citet{calvano2020artificiala}, allowing us to compare our results to theirs irrespective of the absolute profits. The profit gain is the normalized average profit compared with a one-shot Nash Equilibrium price and a monopoly price. In our model with price levels ranging from 0 to 6, 
we take the competitive price level of the one-shot game to be the $(1,1)$ one-shot Nash Equilibrium that leads to $\pi_{c}=3$ profit for both firms. Then, we take the maximum monopoly profit to be $\pi_m=6$, reached at joint prices levels $(3,3)$ and $(4,4)$. Then the profit gain $\Delta_i$ for player $i$ is defined as:

\begin{equation}
    \Delta_i = \frac{\pi(\theta_1, \theta_2)_i - \pi_c}{\pi_m - \pi_c}
\end{equation}

Note that our model allows players to play a price of $0$. This is equivalent to an overly aggressive undercutting below marginal cost, since deviating from a joint price of $(1,1)$ to a price of $0$ is never profitable for either player in our model. Nevertheless, randomized exploration will occasionally lead a $Q$-learner to pick a price of $0$.

\section{Equilibrium and Collusion Phenomena}

In this section, we cover the main results of our paper. We address the questions of competition and collusion in the meta-game with implications for \textit{tacit} collusion. Any parameter profile that is not mutually consistent with best-responding behavior, which generates higher payoffs for both players, in our view, constitutes explicit collusion.

\begin{center}
\emph{Given the parameters of the $Q$-learner on the other side, what parameters should one's own $Q$-learner be equipped with that would generate the highest payoffs for the algorithm designer unilaterally?}
\end{center}

There exist parameter profiles $(\theta_1, \theta_2)$ for ego and alter that are mutual best responses, thus constituting a Nash equilibrium of the meta-game. We call these equilibria the Meta Nash Equilibria (meta-NE), or Meta Nash for short (MN). 

\begin{prop}
    In our game there are pure meta Nash Equilibria. Their 
    pricing behavior is close to the competitive stage-game-Nash.
\end{prop}

We find 
\begin{enumerate}
    \item  two pure symmetric meta-NEs determined by parameters $\alpha=0.12$, $\epsilon=0.0556$ and $\gamma \in \{0,0.11\}$. Pure symmetric meta-NEs lead to aggressively competitive pricing, with a profit gain $\Delta = -0.07$. 
    \item an asymmetric meta-NE with parameters $(\alpha_1, \alpha_2) = (0.01, 1)$, $(\epsilon_1, \epsilon_2) = (0.1667, 0.5)$, and $(\gamma_1, \gamma_2) = (0, 0)$. The asymmetric meta-NE leads to near competitive pricing behavior, with profit gains $\Delta_1 = 0.02$ and $\Delta_2=0.05$.
\end{enumerate}

Profit gains $\Delta\approx0$ imply that profits do not get better than the stage-game-Nash Equilibrium of the Bertrand Duopoly. With the symmetric meta-NE below the stage game Nash, and the asymmetric only slightly above it, if algorithm designers compete in the meta-game, their pricing behavior is bound to be competitive. This leads us to the following question:

\begin{center}
\emph{Given the parameters of both $Q$-learners, what are the combinations that jointly maximize their payoffs?}
\end{center}

Algorithm designers can improve their profits by colluding in the meta-game and co-paremeterizing their $Q$-learners. We get the following

\begin{prop}
    The Pareto frontier of the meta-game consists of asymmetric parameter combinations ($\theta_1 \neq \theta_2$) and their resulting profits are higher than the stage-game-Nash but lower than the stage-game monopoly profits.
\end{prop}

The profit gains $\Delta_1$ and $\Delta_2$ differ, and are distributed accordingly: the maximum profit gain $\Delta$ is $0.56$, the mean profit gain $\Delta$ is $0.26$, and the minimum profit gain $\Delta$ is $-0.23$.  While above the stage-game-Nash, the profits are far below the stage-game monopoly profits. Furthermore, these deviations from meta-NEs are obvious, hence detectable: it suffices that one or both players \textit{increase} their exploration rates ($\epsilon$) and high discount factors, ($\gamma$) and/or learning rates ($\alpha$).

\section{Unilateral Designer Incentives in the Meta-Game}\label{sec:best-repsonses}

\begin{figure}
    \centering
    \includegraphics[width=\linewidth]{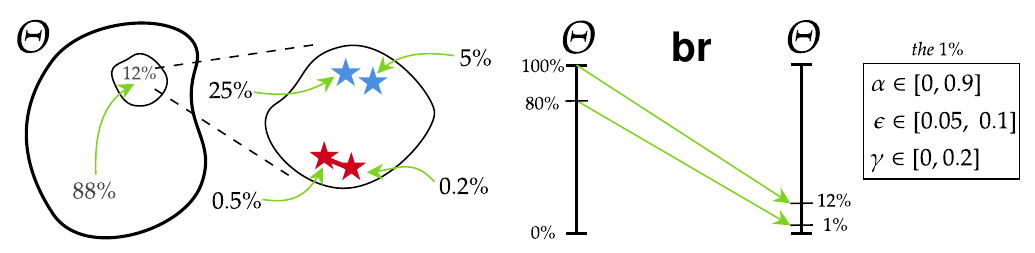}
    \caption{Best responses and relevant ratios. $\Theta$ is the parameter space, \textbf{br} is the best response function, arrows represent best responses, and percentages the percentage of $\Theta$ that have a best response in the same region of $\Theta$. Only about $12\%$ of parameters are best responses, the symmetric meta-NEs are best responeses to $30\%$ of the parameter space, and the asymmetric meta-NEs only $0.07\%$. We have an $80:1$ ratio, where about $80\%$ of the best responses are contained in about $1\%$ of the parameter space. These `top $1\%$' parameters have low exploration rates ($\epsilon$), and low discount factors ($\gamma$).}
    \label{fig:br_ratios}
\end{figure}

Now we can look at the endpoint ratio. First, of the $1$ thousand parameter choices $\theta \in \Theta$ for players (for a total of $1$ million action profiles), a unilateral best response from $\sim 30\%$ of them leads to the symmetric meta-NEs (see \autoref{fig:br_ratios}), while the asymmetric meta-NEs are unilateral best responses to about $0.7\%$ of the parameter space.

\begin{prop}
    Define $BRD: \Theta^2 \rightarrow \Theta^2$ as a function that receives \textit{initial} parameters as input and returns the \textit{end} parameters of best response dynamics. The fraction of starting parameters with end parameters $(\theta^{end}_1, \theta^{end}_2)$ we call the `endpoint ratio' and is defined as:
    $$er(\theta_1, \theta_2) = \frac{1}{|\Theta^2|} \cdot |\{\theta_{1}, \theta_{2} \:|\: BRD(x, y) = (\theta_1, \theta_2), \:\: \forall (x, y) \in \Theta^2 \}|.$$
    Then, the endpoint ratios for the meta-NEs are:
    \begin{enumerate}
        \item symmetric I: $\theta = (0.12, 0.0556, 0)$, $er(\theta, \theta)=0.766$, 
        \item symmetric II: $\theta = (0.12, 0.0556, 0.11)$, $er(\theta, \theta)=0.212$,
        \item asymmetric: $\theta_1 = (0.01, 0.1667, 0), \theta_2= (1, 0.5, 0)$, $er(\theta_1, \theta_2) = 0.022$.
    \end{enumerate}
\end{prop}

\begin{figure}[ht!]
    \centering
    \includegraphics[width=\linewidth]{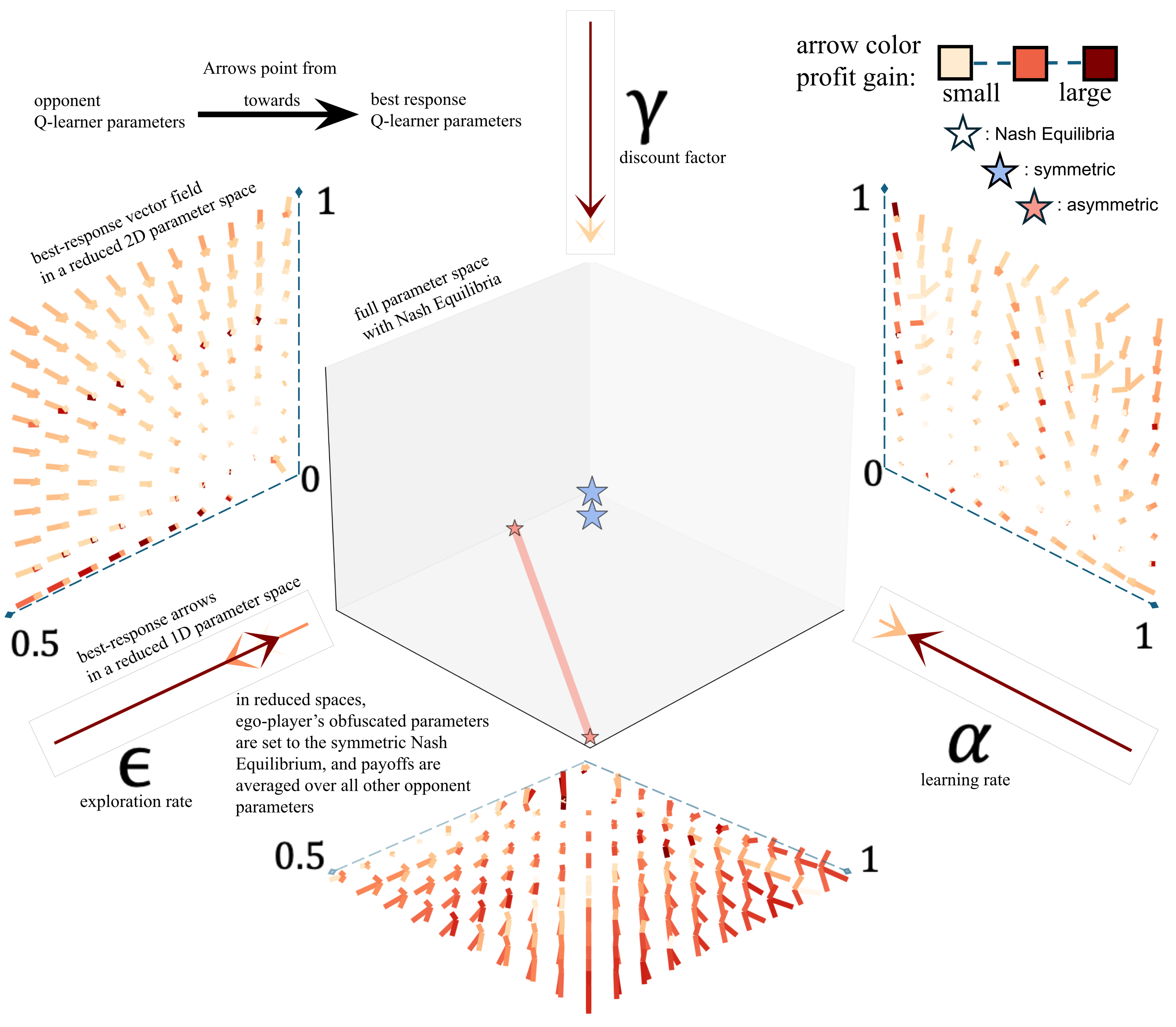}
    \caption{Best Responses in the Meta-Game: this plot encodes best responses as vectors, where the tail starts at the parameters of one player, and the head points to the best response parameters for the other player. The 2D face plots are best responses for payoffs averaged over the third parameter. Similarly, the 1D plots which align with the cube axes average the payoffs over the 2 missing parameters. The figure is explained in greater detail in \autoref{sec:figure_explanation}.}
    \label{fig:best_response_flower}
\end{figure}

\autoref{fig:br_ratios} provides a stylized summary of best responses. The subset of the parameter space which are best responses to at least one other parameter combination is only about $10\%$ of the parameter space, ($127$ of $1000$ parameters to be exact). In \autoref{fig:best_response_flower}, we show the meta-NEs and their relative position in the parameter space.

The best responses are highly concentrated in the neighborhood of meta-NEs. Accordingly, $\sim 80\%$ of best responses lead to $1\%$ of the parameter space, where the values of the discount factor $\gamma$ and the exploration rate $\epsilon$ are small: $\gamma \leq 0.22, 0.0556 \leq \epsilon \leq 0.1111$. It is never a best response to pick a large discount factor. None of the meta-NEs have a discount factor greater than $0.11$. In the 2D parameter fields of \autoref{fig:best_response_flower} where $\gamma$ and $\epsilon$ are present, the direction of the best responses leads to reducing the parameters towards the meta-NEs.

In the `top $1\%$' (see \autoref{fig:br_ratios}) the learning rate $\alpha$ varies most. The variation in $\alpha$ best responses can be explained by the variation in $\gamma$. We can see in \autoref{fig:best_response_flower}, in the 2D parameter field where $\alpha$ and $\gamma$ vary, that the $\alpha$ best responses for higher values of $\gamma$ keep $\alpha$ the same. This shifts when $\gamma$ is very low and close to its meta-NE values, where the $\alpha$ best responses lead to low values of $\alpha$.

\begin{center}
    \textit{Are decayed exploration rates a best response in online algorithmic competition?}
\end{center}

We observe multiple scenarios which we will briefly discuss. 
\begin{scene}
    Alter has an exponentially decaying exploration rate and ego responds with a fixed exploration rate. The payoffs of ego and alter are evaluated online. When alter has high values of the exploration rate it is easy for ego to pick a smaller exploration rate and exploit alter's randomness.

    Fixing alter to have an exploration rate between $0.3 \leq \epsilon \leq 0.5$, the best response exploration rates for ego lie in the range $0.0556 \leq \epsilon \leq 0.1111$ for $93.5\%$ of alter's possible parametrizations.
\end{scene}

\textit{Remark.} Our meta-game suggests that a high exploration rate is rarely going to be a best response. This can be intuitively understood as a high exploration rate leading to random behavior that can be exploited by the opponent and leads to excessive losses. However, our experiments only deal with fixed exploration rates. In practice, it is possible and likely that the optimal exploration strategies involve dynamic and strategic adjustments of the exploration rates. In fact, we assume that practical deployments of reinforcement learning in markets will need to be mostly deterministic, with low exploration rates, most of the time. Then, in precise targeted moments, the algorithms will need to explore to learn more about the market, before returning to mostly deterministic behavior. To show it we run further experiments comparing a decayed exploration rate against a fixed exploration rate in training, and refer the interested reader to our appendix \autoref{sec:fixed-vs-decayed}.

Thus far, we have considered the outcomes of perfectly competitive meta-strategies. If we assume that there are exogenous factors that complicate perfect competition, it may be reasonable to consider the implications of \textit{less than perfect} competition in the meta-game.


\begin{scene}
    \textbf{Ego and alter are both oblivious} players and are clueless about optimal parametrization of their algorithms. Ego and alter randomly pick parameters from $\Theta$. Then
    
    \begin{itemize}
        \item The probability that neither player is profitable is $0.27$,
        while the probability that both players are profitable is $0.25$.
        \item The average profit gain (loss) for both players is $\Delta = -0.16$.
    \end{itemize}
\end{scene}

From the perspective of a single player there is therefore a more than $50\%$ chance of not being profitable when both players are oblivious. This result suggests that, absent knowledgeable algorithm designers, oblivious deployment of a $Q$-learner is hard to justify. 

\begin{scene}
    \textbf{Alter is oblivious}, but \textbf{ego best responds} to the parameters of alter. Alter randomly pick parameters from $\Theta$, and ego picks $\theta_{ego} = br(\theta_{alter})$. Then
    
    \begin{itemize}
        \item The probability that neither player is profitable is $0.15$,
         while the probability only ego is profitable is $0.81$.
        \item The average profit gain for ego is $\Delta=0.13$, and ego's best response improvement is $35\%$ (marginal return over Scenario 1).
    \end{itemize}
\end{scene}

Here we see that a random player is still sufficient to lower the profitability for both, because there are always parameterizations which lead to overly aggressive and ultimately detrimental pricing behaviour. Nonetheless, with a best response ego is able to be profitable more than $80\%$ of the time with a $35\%$ improvement on profits.

\begin{scene}
    \textbf{Ego best responds} to alter, while \textbf{alter stays profitable}. Ego picks $\theta_{ego} = br(\theta_{alter})$, and alter is only guaranteed to pick a parameter $\theta_{alter}$ such that $\Delta_{alter} \geq 0$.

    \begin{itemize}
        \item All combinations are profitable for both players.
        \item The average profit gain when both players are profitable is $\Delta=0.11$, while the average profit gain for ego after best responding is $\Delta=0.19$, which is a $0.07\%$ improvement after best responding.
    \end{itemize}
\end{scene}

When alter is only guaranteed to stay profitable, alter may or may not be picking a parameter which is a best response to ego's. It is interesting to see how just staying profitable is enough to secure rather large profits compared with the meta-NEs. In fact, there is a roughly $35\%$ return over the meta-NEs if both players are able to stay profitable. Interestingly, `staying profitable' may be akin to `conscious parallelism' and not an intent prosecutable for antitrust violation. We will look in detail at `conscious parallelism' by characterizing the Pareto Front of the meta-game in \autoref{sec:collusion}.

\section{Designer Collusion at the Meta-Game Level}\label{sec:collusion}

The literature on algorithmic collusion has focused on price collusion. In this paper, we contribute a new kind of perspective on collusion, which we call algorithm orchestration, representing collusion in the meta-game. Meta-game collusion matters because non-competitive behavior in the meta-game may be the force driving supra-competitive profits. In this section we explain in detail the relationship between price collusion and algorithm orchestration. Given our previous analysis of the Meta Nash Equilibria, we now turn our attention to meta-game collusion.

\begin{figure}
    \centering
    \includegraphics[width=\linewidth]{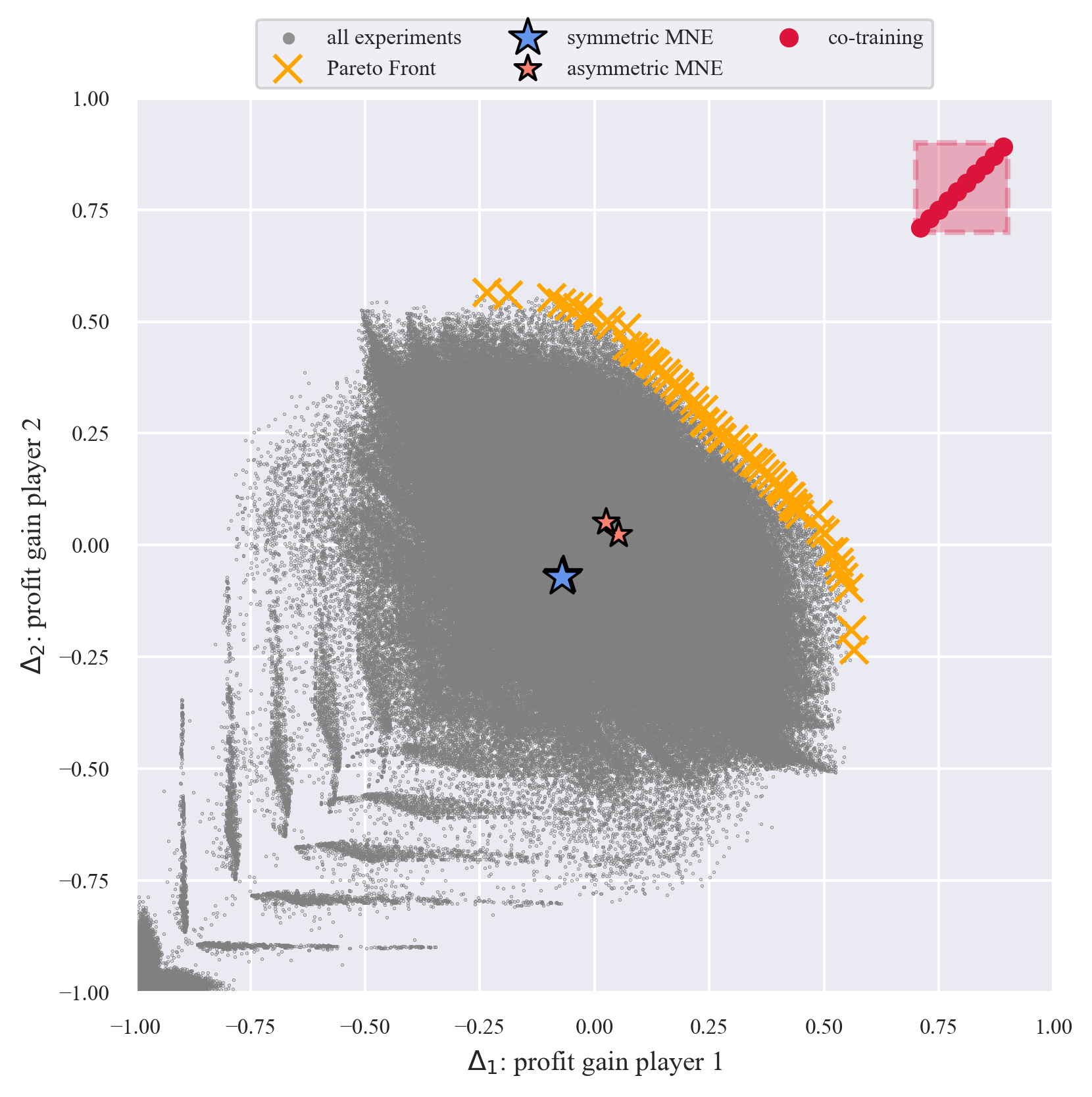}
    \caption{Profit gain comparison between algorithmic design setups. All of our experiments are plotted with tiny gray dots. The meta-NEs are stars (large-blue symmetric, small-red asymmetric), the co-paremetrized Pareto front are crosses (orange) and co-training collusion is red dots enclosed by a box. The co-training collusion values are the range of profit gains achieved in \cite{calvano2020artificiala}.}
    \label{fig:meta_comparisons}
\end{figure}


It is known from the literature that $Q$-learners can achieve highly collusive pricing for profit gains between $0.7$ and $0.9$ \citep{calvano2020artificiala}. However, in our meta-game we achieve an average of $0.25$ on the Pareto front (see \autoref{fig:meta_comparisons}). This discrepancy suggests that $Q$-learning needs joint exploration, coordination in training, and coordination in deployment to learn highly collusive pricing: a form of centralized training and decentralized execution that we call \textit{co-training}. First, algorithm designers need to co-locate, communicate, create a shared training environment, and coordinate their training process. Second, they need to base their model on a representative simulation of demand that matches their deployment conditions. Third, they need to coordinate the deployment of their algorithms in the market. However, co-training is likely to leave many traces of communication that are easy to find and can be used by regulators and antitrust law enforcers to establish the intent of collusion. Therefore, co-training, while highly successful at establishing collusive pricing, is a high effort and high risk strategy for collusion via algorithm orchestration in the meta-game.

\begin{figure}
    \centering
    \includegraphics[width=0.6\linewidth]{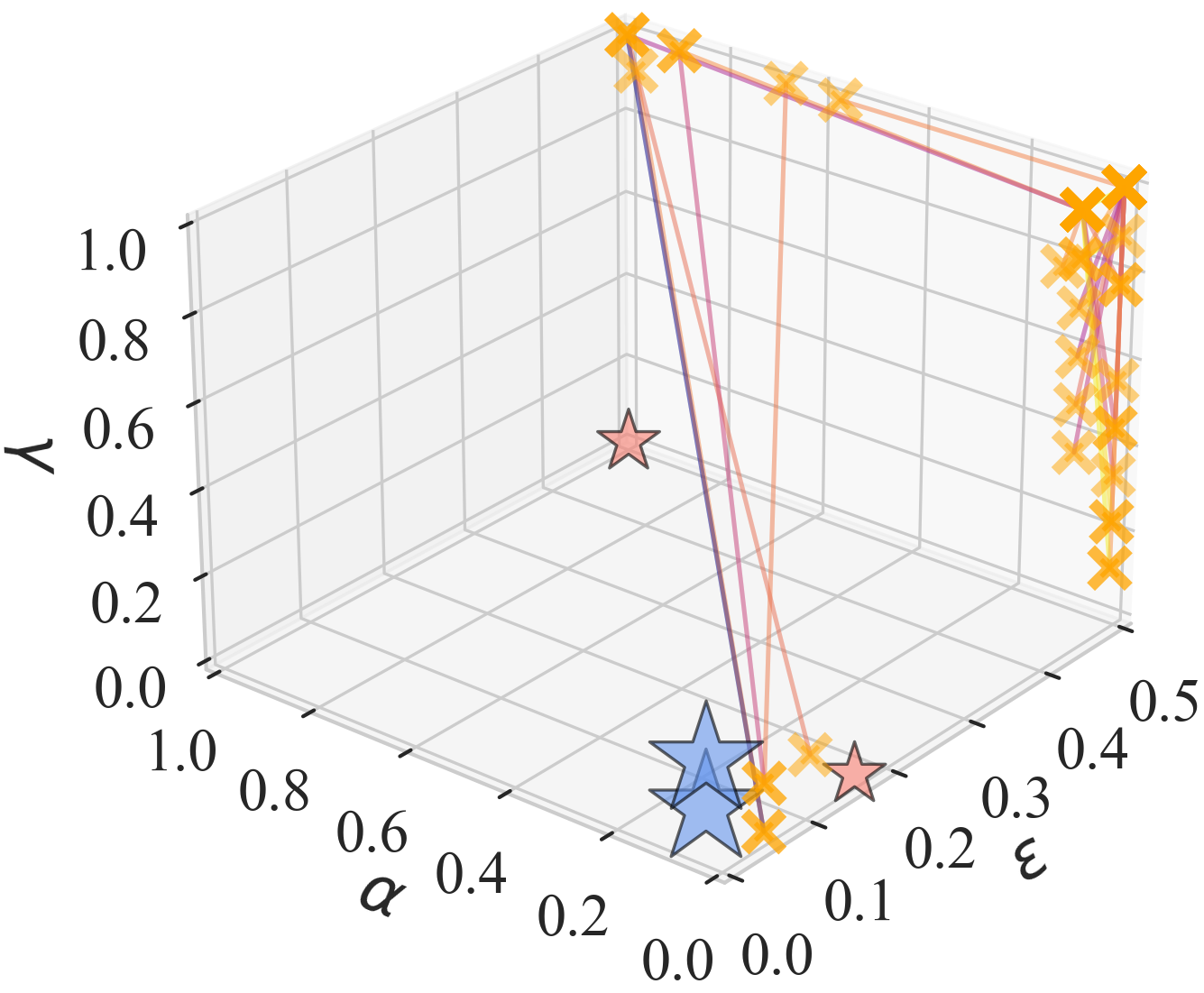}
    \caption{The Pareto Front parameters where both players are profitable $\Delta \geq 1$. The orange crosses connected by lines represent the Pareto optimal combinations, and the meta-NEs are represented as stars. We can identify two kinds of Pareto optimal combinations: a) where one of the parameters is very close to the meta-NEs, and b) where both parameters are far from meta-NEs.}
    \label{fig:pareto-front}
\end{figure}

Unlike co-training, co-parameterization at the Pareto front requires less orchestration and can still result in profit gains. For co-parameterization, algorithm designers only need to communicate their respective learning parameters $(\alpha, \epsilon, \gamma)$ during online learning, obviating the complications of shared training environments. For co-parameterization to be profitable, players can communicate their parameters and then determine responses that will improve their profits without hurting the other's profits. An increased profit gain $\Delta_1$ for player $1$ is followed by a decreased profit gain $\Delta_2$ for player $2$. This relationship can be seen clearly in \autoref{fig:meta_comparisons}. This means that players are not indifferent between the outcomes on the Pareto front. In \autoref{fig:pareto-front} we exclude the Pareto combinations where one of the two players is not profitable, and only look at the combinations where both players achieve a $\Delta \geq 0$.

\begin{prop}
    There are two types of Pareto optima that are profitable for both players ($\Delta_1 \geq 0$ and $\Delta_2 \geq 0$): {\bf Type 1} when one of the two players has parameters in the neighborhood of meta-NEs, and {\bf Type 2} when neither player has parameters close to meta-NEs.
\end{prop}

In Type 1 Pareto optima, one player appears to be maximizing profits. However, the other player is worse off in the co-parameterization, and has a strong incentive to deviate.

\begin{scene}
    Alter picks a high $\alpha > 0.6$ a high $\epsilon > 0.4$ and a high $\gamma > 0.9$. Ego picks a low $\alpha < 0.1$ a low $\epsilon < 0.15$ and a low $\gamma < 0.15$. Ego's parameters $\theta_{ego}$ appear to be a best response. Then

    \begin{itemize}
        \item Ego's profits are between $11\%$ to $46\%$ greater than alter's. 
        \item Alter stands to gain a profit increase between $11\%$ and $13\%$ by unilaterally deviating to better parameters, reducing $\alpha$ and $\gamma$. 
        \item Ego stands to gain less profit by unilaterally deviating to better parameters, between $3\%$ and $10\%$.
    \end{itemize}
\end{scene}

Therefore, a collusive co-parameterization of this sort may be destabilized by the incentives for unilateral deviation. Enforcement of this kind of collusion is likely to require punishments or side-payments. In the meta-game, tacit collusion is hard to sustain without algorithm orchestration. The meta-NEs of the meta-game are a first candidate for successful bilateral profit maximization, and clearly show that price collusion is minimal. The meta-NEs result in aggressively competitive noisy pricing and, as such, are not concerns for tacit collusion. Any tacit collusion requires the establishment of some kind of parameterization that is not a mutual best response.

\section{Conclusions}\label{sec:conclusion}

We have analyzed the strategic incentives of $Q$-learning algorithm designers formulated in a meta-game. We distinguished between algorithmic competition and algorithm orchestration in a repeated Bertrand Duopoly played by $Q$-learners with non-decayed $\epsilon$-greedy policies who condition their actions on the last price of their opponent. The algorithm designers can either parametrize their algorithms competitively, or collusively.

Competition in this meta-game is characterized by its pure and symmetric Nash Equilibria in fixed parameter values, which we call the Meta Nash Equilibria. At the Meta Nash Equilibria we find aggressively competitive and noisy pricing behavior. We also consider slight relaxations of perfect competition in the meta-game to find slightly supra-competitive pricing, but with large incentives for unilateral deviation.   

Collusion in the meta-game by contrast is defined by co-training and co-parameterization. Co-training is the process of jointly training and deploying algorithms, and requires extensive orchestration of algorithms. Co-training results in the kinds of collusive pricing behaviors that can achieve close to monopoly prices \citep{calvano2020artificiala} and Edgeworth cycles \citep{klein2021autonomousa}. Co-parameterization is the process of pursuing a joint profit objective and can be achieved with as little as the sharing of learning parameters between designers. Co-parameterization leads to pricing that is very different from the previously identified collusive behavior in the literature of focal price collusion and Edgeworth cycles that are obtained with co-training.

In this concludin section, we shall first outline how one might be able to test for algorithm orchestration before discussing limitations and avenues for further research.

\subsection{A test of tacit collusion}

\begin{proposal}
    We can test the Meta Nash hypothesis: given pricing patterns of two players we detect whether or not the prices are generated by $Q$-learners parametrized at the Nash Equilibrium of the meta-game.
\end{proposal}

\begin{algorithm}[t]
\SetAlgoNoLine
\KwIn{Two pricing streams $X_1,X_2$ generated from unknown parameters $\theta_1, \theta_2$.}
\KwOut{Test of Meta Nash Hypothesis $\mathcal{H}_{MN}$: whether $\theta_1, \theta_2$ are Nash Equilibrium.}

\eIf{the average of $X_1$ and $X_2$ is close to bottom of the price range}{
    fail to reject $\mathcal{H}_{MN}$
}{
    \eIf{$X_1$ or $X_2$ are noisy}{
        \eIf{$X_1$ and $X_2$ are symmetric}{
            Communication is necessary for symmetric supra-competitive pricing \\
            $\rightarrow$ Reject $\mathcal{H}_{MN}$ in favor of co-parameterization with communication.
        }{
            Communication is not necessary for asymmetric supra-competitive pricing \\
            $\rightarrow$  Reject $\mathcal{H}_{MN}$ in favor of co-parameterization via `conscious parallelism'.
        }
    }{
    Communication is necessary for noise-free supra-competitive pricing \\
    $\rightarrow$ Reject $\mathcal{H}_{MN}$ in favor of co-training.
    }
}

\caption{Meta-game mutual best response detection}
\label{alg:one}
\end{algorithm}

We provide \autoref{alg:one}, a step-by-step schematic to test the Meta Nash hypothesis $\mathcal{H}_{MN}$: that players have picked mutual best response parameters in the meta-game. The differences between Meta Nash pricing, Pareto front pricing and decayed exploration pricing are substantial. This suggests that statistical tests on the market prices may be enough to determine whether or not players are at the Nash Equilibrium or are colluding in the meta-game. 

We start by checking whether two pricing streams $X_1$ and $X_2$ are frequently close to the minimum of the price range. If they are, we can not exclude competitive play in the meta-game. If the pricing streams are frequently higher than the minimum, we proceed to check whether they are noisy. By noisy we mean that there is stochasticity in the price changes that can only be explained by an exploration policy. It may be possible to detect stochasticity driven by an exploration policy if it is uncorrelated with exogenous demand shocks. If the price streams do not have stochasticity uncorrelated with exogenous demand shocks, then we reject $\mathcal{H}_{MN}$ in favor of co-training, because it is only via co-training that we know for supra-competitive and noise-free pricing to be possible, with focal prices and Edgeworth cycles.

If both price streams are noisy, then we check whether or not they are symmetric. By symmetric, we mean that we can find the same statistical regularities of one price stream in the other. We take symmetric statistical regularities as an indication of symmetric parameterization in the meta-game, co-parameterization. If the co-parameterization is symmetric and supra-competitive, we argue that some communication was necessary to establish the coordination. If instead the price streams are supra-competitive but do not show symmetric statistical regularities, we have an asymmetric form of co-parameterization where communication may not be necessary, and a joint profit objective may be enough.

\subsection{Limitations and Future Work}

A Bertrand Duopoly played by $Q$-learning is a useful model to reason about price competition, but an unlikely scenario to encounter in the real-world. The main contribution of this paper is the framework by which we judge competitive design choices from collusive design choices for competing algorithms. This may be helpful in defining a category of `prohibited algorithms' \citep{harrington2018developingb}. However, there are possibly infinitely many classes of algorithms that would need to be evaluated computationally against each other to determine this category \textit{ante facto}. Instead, we advocate for an \textit{ad hoc} evaluation of algorithm design competitiveness. Therefore, the greatest limitation at this present moment is our lack of knowledge of the pricing algorithms that companies are \textit{actually using}.

Furthermore, while our results suggest that there exists a pure Nash Equilibrium in the meta-game that we analyze, it is not clear what this implies for general meta-games. It is likely that the meta-game equilibrium structure depends on many factors that we do not vary in this experiment. Certainly, changing the underlying algorithms of the meta-game will influence the outcomes. Similarly, it is likely that the aggregation of rewards during repeated play also has an effect. We conjecture that a meta-game played by symmetric algorithm designers in an underlying repeated game that is also symmetric will yield meta-equilibria that are symmetric, but we have to leave further proof and evidence to future work.

In future work, we are interested to introduce demand uncertainty to the underlying game. To what extent is algorithmic collusion, which boils down to a coordination problem, disrupted by such noise? Additionally, we encourage future work to evaluate algorithms \textit{online} (or cumulatively) rather than \textit{offline} (or in the limit), because evaluation in the limit rests on weakly supported assumptions that exploration is costless. In other words, it assumes that price discovery does not require interaction with a market. This is highly unlikely to be the case, and our results suggest that this assumption yields misleading conclusions.

Furthermore, it would seem that when exploration is costless and learning is evaluated in the limit, the optimal behavior of players is to learn as much as possible about the pricing strategy of the opponent. However, what does one player learn about the pricing strategy of the other player, when both players are cost-less-ly exploring each others pricing strategy? They would learn each others \textit{training} strategies and not their \textit{pricing} strategies. In such a setup, the players do not learn to play against the opponents they will not face during deployment. Therefore, we argue for the analysis of \textit{online} learning methods instead of \textit{offline}, \textit{train-then-deploy}, and \textit{centralized training decentralized execution} style approaches that are popular in orchestrated systems.

We conclude this paper by pointint out that finance, housing, telecommunications and platforms in general are domains where pricing algorithms  are increasingly widespread. This may turn into a major problem, particularly as a growing literature has shown that they are able to learn to collude. We argue that in order to regulate the phenomenon meaningfully we must meaningfully differentiate between algorithm competition, when designers choose competitive parameters, and algorithm orchestration, when designers coordinate their parameters to collusive outcomes.
Our paper indicates that supra-competitive prices between $Q$-learning algorithms can be achieved with algorithm orchestration. Pertinently, we show that the Nash Equilibrium of the meta-game of parameter picking leads to noisy competitive prices. These algorithms generate profits which are slightly supra-competitive but generally beneficial for the welfare of consumers. If we extend the definition of non-collusion of \citet{hartline2024regulation}, that pricing behavior should not be considered collusive if it is approximately best responding to the opponent, then we \textit{cannot} consider two players that play the meta-game close to Nash to be colluding with one another. On the other hand, any price pattern similar to steady focal prices or clean Edgeworth cycles may be evidence of collusion by algorithm designers, and regulators should start by out-ruling algorithm designer collusion; i.e. algorithms that are being checked must be shown to be parametrized to maximize payoffs unilaterally, or at least be tuned to that goal.

\backmatter

\bmhead{Availability of data and materials}

The simulation code and data can be made available upon request.

\bmhead{Conflict of interests}

The authors declare that they have no competing interests

\bmhead{Funding}

CC received funding from the ERC CoCi grant no. 833168. HN was supported by the SNSF Eccellenza Grant `Markets and Norms.' FF acknowledges the support of National Science Centre, Poland, grant nr. 2023/51/B/HS4/01343.

\bmhead{Authors' contributions}

CC is the lead author of the manuscript. He designed and implemented the model, ran the experiments and analyzed the data. 
HN contributed to the design of model and experiments.
FF and SR helped with formalization and design of the simulation model. 

\bmhead{Acknowledgements}

We gratefully acknowledge helpful conversations with Helmut Bölcskei, Arthur Dolgopolov, Dirk Helbing, Timo Klein, Janusz Meylahn, Nicholas Eschenbaum and with participants of the Concurrences Global Conference on Antitrust (NY, 2025) Joseph Harrington, Dennis Carlton, Michael Jo, and Gwendolyn Cooley.

\newpage

\begin{appendices}

\section{Collusion Literature Table}

Research on the broader topic of Algorithmic Collusion/Cooperation in a variety of games. `Paper' references are in the bibliography. We organized the literature according to several dimensions of design choices and results. In terms of `Game,' we differentiate between Prisoner's Dilemma (PD), Cournot Oligopoly (CO), Price Competition (PCG), Storage Capacity (SCG), Population Game (PG), and Braess Paradox (BP). In terms of learning rules of the `Agents,' we have Q-learning (Q), Tit-for-Tat (TfT), Upper Confidence Bound (UCB), and Particle Swarm Optimization (PSO). In terms of `Exploration' approach, we have $\epsilon$-greedy ($\epsilon$-g), and Boltzmann exploration ($T\degree$). We also distinguish papers by whether they allow agent asymmetry (asym\_) and what kind of evaluation type (eval\_) they apply. In terms of results, we describe their main result regarding collusion, and whether the algorithm designer picks parameters at Nash Equilibrium (NE) of an associated meta-game. Note that *** indicates our present paper.

\begin{longtable}{l|l|l|l|l|l|l}
\caption{Wider Algorithmic Collusion Literature}
\label{tab:broader_literature}\\
\toprule
             Paper &                            Game &                  Agents &                  exploration & asym\_ &                       eval\_ & insights \\
\midrule
\endfirsthead
\midrule
\endhead
\midrule
\multicolumn{7}{r}{{Continued on next page}} \\
\midrule
\endfoot

\bottomrule
\endlastfoot
   \citetalias{sandholm1996multiagenta} &              PD & Q, TfT &           $T\degree$, decay &                 yes &                         limit & $T\degree$ increases collusion \\
\citetalias{waltman2008qlearning} &               CO &                     Q &           $T\degree$, decay &                  no &                         limit & $\alpha$ increases collusion \\
\citetalias{wunder2010classesb} &              PD &                     Q &     $\epsilon$-g, fixed &                  no &                       online & oscillating collusion \\
\citetalias{hansen2021frontiers} &               pc &                            UCB &                 UCB, decay &                  no &                         limit & signal-to-noise \\
\citetalias{sanchez-cartas2022artificial} &               pc &             Q, PSO & $\epsilon$-g, decay &                 yes &                         limit & Q more collusive than PSO \\
\citetalias{abada2023artificial} &                sc &                     Q &   $\epsilon$-g, decay &                  no &                         limit & `rational' exploration  \\
\citetalias{banchio2023adaptive} &              PD &                     Q &     $\epsilon$-g, fixed &                  no &                         limit & collusion by coupling \\
\citetalias{leonardos2023catastropheb} &                 PG &                     Q &                    $T\degree$ &                  no &                         limit & equilibrium selection \\
\citetalias{hartline2024regulation} &               pc &                            n.a. &                          n.a. &                 n.a. &                       online & price auditing method \\
\citetalias{carissimo2024counterintuitive} & BP & Q & $\epsilon$-g, fixed & no & online & cycles with fixed $\epsilon$ \\
C25 & BP & Q & $\epsilon$-g, fixed & yes & online & Edgeworth cycles \\ \hline
\end{longtable}

\section{Best Responses}

\subsection{Explaining \autoref{fig:best_response_flower}}\label{sec:figure_explanation}

\autoref{fig:best_response_flower} is the main visualization of our results for the meta-game. Given the high dimensionality of the parameter space we show the results in a compressed representation that conveys a wealth of information. The figure represents the best response relationship between all parameter profiles in a 3 dimensional space. This figure has three types of plots aligned with the main axes: the cube at the center, the line plots aligned with the interior axes, and the 2D plots aligned with the faces of the cube. 

The cube plot at the center of \autoref{fig:best_response_flower} is an encoding of the entire parameter space. Each point in the cube is a valid parameter setting for a single $Q$-learner.

The line plots are aligned with the main axes of the cube. We take a reference parameter e.g. $\alpha$ and assume that ego's other parameters e.g. $\epsilon, \gamma$ follow a symmetric Meta Nash strategy ($\alpha=0.12$, $\epsilon=0.0556$, $\gamma=0.11$). Then, we compute the best response reference parameter of ego given alter's reference parameter, and keeping ego's other parameters fixed. We take the average over alter's other parameters to determine the payoff. 

The surface plots are aligned with the faces of the cube. We take two reference parameters e.g. $\alpha, \gamma$ and set ego's other parameter e.g. $\epsilon$ to a Meta Nash strategy ($\alpha=0.12$, $\epsilon=0.0556$, $\gamma=0.11$). Then, we compute the best response of ego assuming that ego can vary both reference parameters while taking the uniform weighted average profit over alter's parameters as the payoff. 

These reduced parameter plots can be interpreted as a form of decision making under uncertainty. Given that the reference player only knows the reference parameter of the opponent, we assume the other parameter settings are equiprobable. Then, we assume that the reference player considers the best responses in a reduced space of the parameters, setting their other parameters to a `known' dominant strategy, the Meta Nash values of those parameters. We note that other distributions over the unknown parameters could be considered, and that these may change the reduced equilibria. These face plots should only be used to develop intuition about the best response relationships in the entire parameter space.

The arrows in the space start at the parameters of alter and point to the best response parameters of ego. All arrows are colored according to the payoff gain obtained by deviating to the best response, assuming that the starting point was the symmetric parameter profile at the tail of the arrow.

\subsection{Fixed exploration rate vs decayed exploration rate}\label{sec:fixed-vs-decayed}

In this section, we compare decayed exploration against fixed exploration rates. This analysis is relevant, because decayed exploration policies constitute the most common design choice in the literature. Crucially, we analyze the online payoffs achieved by the algorithms during training, instead of the profits achieved after convergence. 

Under decayed exploration, $Q$-learners have an $\epsilon$-greedy policy as defined in \autoref{sec:methods}, and the exploration rate $\epsilon$ is decayed exponentially with the form $\epsilon=\exp(-\beta t)$, $\beta>0$ during some initial phase.
Note that the exploration rate is the only parameter which is dynamically adjusted during learning in this model. Decay of exploration naturally has a large effect on the learning dynamics and convergence of the algorithms. Indeed exploration has also been identified in the literature as a central driver of collusive behavior, particularly when it is random and leads to `non-rational' prices \cite{abada2023artificial}. Herein we address the following question of our meta-game: 

\begin{center}
    \textit{Is it a rational choice to decay the exploration rate against an online opponent?}
\end{center}

Our main argument in this section is that a fixed exploration rate is better than an exploration rate which decays to zero. We substantiate this claim with theoretical and empirical arguments.

\begin{figure}
    \centering
    \includegraphics[width=\linewidth]{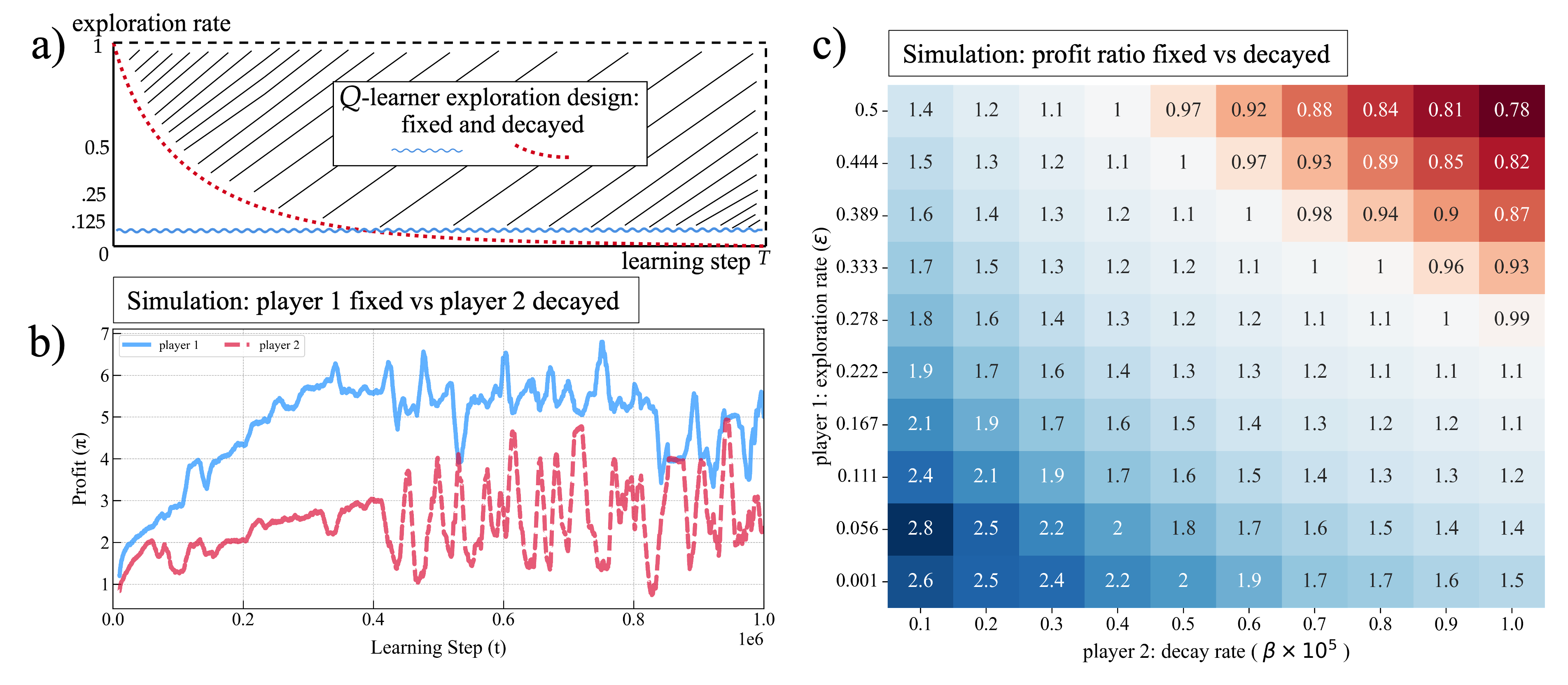}
    \caption{Comparison of fixed exploration and decayed exploration: Player 1 has parameters $\alpha=0.12, \epsilon=0.27, \gamma=0.22$, while Player 2 has parameters $\alpha=0.01, \gamma=0.99$ and $\epsilon$ is decayed from 1 to 0 with decay parameter $\beta=0.4*10^{-5}$.}
    \label{fig:fix-decay-comparison}
\end{figure}

We start with the observation that a $Q$-learner in a Bertrand Duopoly is a non-stationary environment for the opponent. Therefore, $Q$-learning is not guaranteed to converge against another $Q$-learner. Then, we note that the non-stationary dynamics of a $Q$-learner are due to the updating of the $q$-values. The $q$-values represent a pricing strategy, and a $Q$-learning algorithm moves through the space of pricing strategies such that the pricing strategy changes. This time and reward dependent change creates the non-stationary dynamics for the opponent. However, if the opponent has converged and stops changing, they become a stationary MDP.

If an opponent has converged, we can optimally learn to play against them with $Q$-learning. If, on the other hand, the opponent has not yet converged, we should make sure to keep exploring so that we can adapt to their non-stationary dynamics.
Strategically following our claim, both players do not wish to converge against the other before the other player has converged against them. Instead, players wish to balance their exploration against the opponent to adapt to the opponent's $q$-values updates. Moreover, considering the limiting behavior (for very small exploration rate) agents meets inductive bias, where small randometa-NEss hinder learning. Thus, it may be very difficult to move away from local optima when it requires large deviation from the current strategy.

In \autoref{fig:fix-decay-comparison} we compare two $Q$-learners, one with a fixed-exploration rate and the other with a decayed exploration rate. The fixed player (blue) has a constant exploration rate ($\alpha=0.12, \gamma=0.22$) while the decayed player (red) has an exploration rate $\epsilon = \exp (-\beta t)$  ($\alpha=0.01$, $\gamma=0.99$). The second player's parameters  are in a range of highly collusive parameters following \cite{calvano2020artificiala}, while the first players parameters are in a range which we found to dominate most other parameter combinations, which we discuss in detail in the next section. Subfigure a) graphically illustrates the difference between the decayed and fixed exploration rates. In subfigure b) we see a single simulation with time-averaged profits of both players. The fixed exploration player consistently achieves a higher profit than the decayed exploration player. The specific parameters of this simulation are picked with reference to the results of subfigure c) ($\epsilon=0.001, \beta=10^{-5}$).

In \autoref{fig:fix-decay-comparison} subfigure c) is a heatmap of an ablation study of exploration parameters, for both the fixed player (vertical axis) and the decayed player (horizontal axis). The displayed values reflect the profit ratio, the profit of the fixed player divided by the profit of the decayed player, in a simulation of $1$ million learning steps, with each simulation repeated 40 times, and the average profit averaged over all repetitions. Our range of decay rates overlaps with the decay rates tested in \cite{calvano2020artificiala}. Slower decay rates which lead to more exploration, are easier to exploit for the fixed player. Both players benefit from reducing their exploration parameter against the opponent. We find that the fixed player can always achieve higher profits than the decayed player by picking a strategic exploration rate.



\end{appendices}


\bibliography{references}

@article{dolgopolov2024reinforcement,
  title={Reinforcement learning in a prisoner's dilemma},
  author={Dolgopolov, Arthur},
  journal={Games and Economic Behavior},
  volume={144},
  pages={84--103},
  year={2024},
  publisher={Elsevier}
}

@article{abada2023artificial,
  title = {Artificial {{Intelligence}}: {{Can Seemingly Collusive Outcomes Be Avoided}}?},
  shorttitle = {Artificial {{Intelligence}}},
  author = {Abada, Ibrahim and Lambin, Xavier},
  year = {2023},
  month = sep,
  journal = {Management Science},
  volume = {69},
  number = {9},
  pages = {5042--5065},
  issn = {0025-1909, 1526-5501},
  doi = {10.1287/mnsc.2022.4623},
  urldate = {2024-04-24},
  langid = {english}
}

@article{asker2022artificial,
  title = {Artificial {{Intelligence}}, {{Algorithm Design}}, and {{Pricing}}},
  author = {Asker, John and Fershtman, Chaim and Pakes, Ariel},
  year = {2022},
  month = may,
  journal = {AEA Papers and Proceedings},
  volume = {112},
  pages = {452--456},
  issn = {2574-0768, 2574-0776},
  doi = {10.1257/pandp.20221059},
  urldate = {2024-04-24},
  langid = {english}
}

@article{bambauer2018algorithm,
  title = {The {{Algorithm Game}}},
  author = {Bambauer, Jane and Zarsky, Tal},
  year = {2018/2019},
  journal = {Notre Dame Law Review},
  volume = {94},
  number = {1},
  pages = {1--48},
  urldate = {2024-11-26},
  langid = {english}
}

@inproceedings{banchio2023adaptive,
  title = {Adaptive {{Algorithms}} and {{Collusion}} via {{Coupling}}},
  booktitle = {Proceedings of the 24th {{ACM Conference}} on {{Economics}} and {{Computation}}},
  author = {Banchio, Martino and Mantegazza, Giacomo},
  year = {2023},
  month = jul,
  pages = {208--208},
  publisher = {ACM},
  address = {London United Kingdom},
  doi = {10.1145/3580507.3597726},
  urldate = {2025-01-17},
  isbn = {979-8-4007-0104-7},
  langid = {english}
}

@article{calvano2020artificiala,
  title = {Artificial {{Intelligence}}, {{Algorithmic Pricing}}, and {{Collusion}}},
  author = {Calvano, Emilio and Calzolari, Giacomo and Denicol{\`o}, Vincenzo and Pastorello, Sergio},
  year = {2020},
  journal = {The American Economic Review},
  volume = {110},
  number = {10},
  eprint = {26966472},
  eprinttype = {jstor},
  pages = {3267--3297},
  publisher = {American Economic Association},
  issn = {0002-8282},
  urldate = {2024-11-21}
}

@article{carissimo2024counterintuitive,
  title = {Counter-{{Intuitive Effects}} of {{Q-Learning Exploration}} in a {{Congestion Dilemma}}},
  author = {Carissimo, Cesare},
  year = {2024},
  journal = {IEEE Access},
  volume = {12},
  pages = {15984--15996},
  issn = {2169-3536},
  doi = {10.1109/ACCESS.2024.3358608},
  urldate = {2025-01-22}
}

@misc{denboer2024artificial,
  type = {{{SSRN Scholarly Paper}}},
  title = {Artificial {{Collusion}}: {{Examining Supracompetitive Pricing}} by {{Q-Learning Algorithms}}},
  shorttitle = {Artificial {{Collusion}}},
  author = {{den Boer}, Arnoud V. and Meylahn, Janusz M. and Schinkel, Maarten Pieter},
  year = {2024},
  month = nov,
  number = {4213600},
  eprint = {4213600},
  publisher = {Social Science Research Network},
  address = {Rochester, NY},
  doi = {10.2139/ssrn.4213600},
  urldate = {2025-01-17},
  archiveprefix = {Social Science Research Network},
  langid = {english}
}

@article{eschenbaum2021robust,
  title = {Robust {{Algorithmic Collusion}}},
  author = {Eschenbaum, Nicolas and Mellgren, Filip and Zahn, Philipp},
  year = {2021},
  langid = {english}
}

@article{hansen2021frontiers,
  title = {Frontiers: {{Algorithmic Collusion}}: {{Supra-competitive Prices}} via {{Independent Algorithms}}},
  shorttitle = {Frontiers},
  author = {Hansen, Karsten T. and Misra, Kanishka and Pai, Mallesh M.},
  year = {2021},
  month = jan,
  journal = {Marketing Science},
  volume = {40},
  number = {1},
  pages = {1--12},
  publisher = {INFORMS},
  issn = {0732-2399},
  doi = {10.1287/mksc.2020.1276},
  urldate = {2024-04-24}
}

@article{harrington2018developingb,
  title = {{{Developing Competition Law For Collusion by Autonomous Artificial Agents}}{\dag}},
  author = {Harrington, Joseph E},
  year = {2018},
  month = sep,
  journal = {Journal of Competition Law \& Economics},
  volume = {14},
  number = {3},
  pages = {331--363},
  issn = {1744-6414, 1744-6422},
  doi = {10.1093/joclec/nhy016},
  urldate = {2025-01-31},
  copyright = {https://academic.oup.com/journals/pages/open\_access/funder\_policies/chorus/standard\_publication\_model},
  langid = {english}
}

@inproceedings{hartline2024regulation,
  title = {Regulation of {{Algorithmic Collusion}}},
  booktitle = {Proceedings of the {{Symposium}} on {{Computer Science}} and {{Law}}},
  author = {Hartline, Jason D. and Long, Sheng and Zhang, Chenhao},
  year = {2024},
  month = mar,
  pages = {98--108},
  publisher = {ACM},
  address = {Boston MA USA},
  doi = {10.1145/3614407.3643706},
  urldate = {2025-01-17},
  isbn = {979-8-4007-0333-1},
  langid = {english}
}

@article{klein2021autonomousa,
  title = {Autonomous Algorithmic Collusion: {{Q-learning}} under Sequential Pricing},
  shorttitle = {Autonomous Algorithmic Collusion},
  author = {Klein, Timo},
  year = {2021},
  journal = {The RAND Journal of Economics},
  volume = {52},
  number = {3},
  pages = {538--558},
  issn = {1756-2171},
  doi = {10.1111/1756-2171.12383},
  urldate = {2025-01-08},
  langid = {english}
}

@article{leonardos2023catastropheb,
  title = {Catastrophe by {{Design}} in {{Population Games}}: {{A Mechanism}} to {{Destabilize Inefficient Locked-in Technologies}}},
  shorttitle = {Catastrophe by {{Design}} in {{Population Games}}},
  author = {Leonardos, Stefanos and Sakos, Joseph and Courcoubetis, Costas and Piliouras, Georgios},
  year = {2023},
  month = jun,
  journal = {ACM Trans. Econ. Comput.},
  volume = {11},
  number = {1-2},
  pages = {1:1--1:36},
  issn = {2167-8375},
  doi = {10.1145/3583782},
  urldate = {2025-01-22}
}

@article{sanchez-cartas2022artificial,
  title = {Artificial {{Intelligence}}, {{Algorithmic Competition}} and {{Market Structures}}},
  author = {{Sanchez-Cartas}, J. Manuel and Katsamakas, Evangelos},
  year = {2022},
  journal = {IEEE Access},
  volume = {10},
  pages = {10575--10584},
  issn = {2169-3536},
  doi = {10.1109/ACCESS.2022.3144390},
  urldate = {2025-01-17}
}

@article{sandholm1996multiagenta,
  title = {Multiagent Reinforcement Learning in the {{Iterated Prisoner}}'s {{Dilemma}}},
  author = {Sandholm, Tuomas W. and Crites, Robert H.},
  year = {1996},
  month = jan,
  journal = {Biosystems},
  volume = {37},
  number = {1-2},
  pages = {147--166},
  issn = {03032647},
  doi = {10.1016/0303-2647(95)01551-5},
  urldate = {2025-01-22},
  copyright = {https://www.elsevier.com/tdm/userlicense/1.0/},
  langid = {english}
}

@article{schwalbe2018algorithms,
  title = {{{Algorithms}}, {{Machine Learning}}, {{and Collusion}}},
  author = {Schwalbe, Ulrich},
  year = {2018},
  month = dec,
  journal = {Journal of Competition Law \& Economics},
  volume = {14},
  number = {4},
  pages = {568--607},
  issn = {1744-6414, 1744-6422},
  doi = {10.1093/joclec/nhz004},
  urldate = {2025-01-31},
  copyright = {https://academic.oup.com/journals/pages/open\_access/funder\_policies/chorus/standard\_publication\_model},
  langid = {english}
}

@article{waltman2008qlearning,
  title = {Q-Learning Agents in a {{Cournot}} Oligopoly Model},
  author = {Waltman, Ludo and Kaymak, Uzay},
  year = {2008},
  month = oct,
  journal = {Journal of Economic Dynamics and Control},
  volume = {32},
  number = {10},
  pages = {3275--3293},
  issn = {01651889},
  doi = {10.1016/j.jedc.2008.01.003},
  urldate = {2025-01-22},
  copyright = {https://www.elsevier.com/tdm/userlicense/1.0/},
  langid = {english}
}

@article{watkins1992q,
  title={Q-learning},
  author={Watkins, Christopher JCH and Dayan, Peter},
  journal={Machine learning},
  volume={8},
  pages={279--292},
  year={1992},
  publisher={Springer}
}

@inproceedings{wunder2010classes,
  title={Classes of multiagent q-learning dynamics with epsilon-greedy exploration},
  author={Wunder, Michael and Littman, Michael L and Babes, Monica},
  booktitle={Proceedings of the 27th International Conference on Machine Learning (ICML-10)},
  pages={1167--1174},
  year={2010}
}

@inproceedings{wunder2010classesb,
  title = {Classes of {{Multiagent Q-learning Dynamics}}  with $\epsilon$-Greedy {{Exploration}}},
  author = {Wunder, Michael and Littman, Michael and Babes, Monica},
 booktitle={Proceedings of the 27th International Conference on Machine Learning (ICML-10)},
  pages={1167--1174},
  year = {2010},
  langid = {english}
}

@article{brown2023competition,
  title={Competition in pricing algorithms},
  author={Brown, Zach Y and MacKay, Alexander},
  journal={American Economic Journal: Microeconomics},
  volume={15},
  number={2},
  pages={109--156},
  year={2023},
  publisher={American Economic Association 2014 Broadway, Suite 305, Nashville, TN 37203-2425}
}

@article{salcedo2015pricing,
  title={Pricing algorithms and tacit collusion},
  author={Salcedo, Bruno},
  journal={Manuscript, Pennsylvania State University},
  year={2015}
}

@article{lamba2022pricing,
  title={Pricing with algorithms},
  author={Lamba, Rohit and Zhuk, Sergey},
  journal={arXiv preprint arXiv:2205.04661},
  year={2022}
}

@article{harrington2025hub,
  title={Hub-and-spoke collusion with a third-party pricing algorithm},
  author={Harrington Jr, Joseph E},
  journal={Available at SSRN 5010894},
  year={2025}
}

\end{document}